# Controllably asymmetric beam splitting via gap-induced diffraction channel transition in dual-layer binary metagratings

Yangyang Fu[1*], Jiaqi Tao[1], Ailing Song[3], Youwen Liu[1], and Yadong Xu[2*]

[1]*College of Science, Nanjing University of Aeronautics and Astronautics, Nanjing 211106, China*
[2]*School of Physical Science and Technology, Soochow University, Suzhou 215006, China*
[3]*School of Mechanical Engineering, Xi'an Jiaotong University, Xi'an, Shaanxi 710049, Chinaa*
*Email: yyfu@nuaa.edu.cn, ydxu@suda.edu.cn

In this work, we designed and studied a feasible dual-layer binary metagrating, which can realize controllable asymmetric transmission and beam splitting with nearly perfect performance. Owing to ingenious geometry configuration, only one meta-atom is required to design for the metagrating system. By simply controlling air gap between dual-layer metagratings, high-efficiency beam splitting can be well switched from asymmetric transmission to symmetric transmission. The working principle lies on gap-induced diffraction channel transition for incident waves from opposite directions. The asymmetric/symmetric transmission can work in a certain frequency band and a wide incident range. Compared with previous methods using acoustic metasurfaces, our approach has the advantages of simple design and tunable property and shows promise for applications in wavefront manipulation, noise control and acoustic diode.

## 1. Introduction

Freely controlling wave propagation with high-efficiency is strongly desired in wave communities, and recent development of artificial metamaterials [1, 2] offers an outstanding way for wave manipulation with unconventional properties not found in nature. As a 2D counterpart of metamaterials, metasurfaces [3, 4] have drawn great interest of research, as they opened up ways for controlling waves with more advantages, such as ultrathin thickness, planar configuration and simpler fabrication. In acoustic field, by designing various meta-structures with special profiles, acoustic metasurfaces (AMs) [5] have been employed to realize numerous applications, including wavefront manipulation [6-8], holography rendering [9] and perfect absorber [10]. As one of most important applications, asymmetric transmission (AT), enabling one-way wave propagation with highly different transmission efficiency, has been widely explored in both passive and active systems [11-15], yet suffering from respective drawbacks, e.g., bulk structure, implement complexity and low conversion efficiency. By means of ultrathin thickness and planar geometry, AMs have been demonstrated as a good candidate for achieving AT [16-22], such as AMs in waveguides [16, 17], AMs integrated with near-zero index medium [18] and lossy AMs [19, 20]. To obtain smooth wavefront, AMs are required to provide continuous phase shifts covering $2\pi$, which is commonly discretized and realized by many unit cells for high resolution. Therefore, these AMs for AT are generally composed of multiple unit cells, which not only adds the design complexities but also poses a challenge for sample fabrication in higher frequencies [23]. Meanwhile, more unit cells in AMs can bring more sound absorption due to multiple reflection effect [24, 25], which may reduce the performance of AT. Therefore, AMs with simplified design is extremely desirable to control wave propagation, in particular, to realize AT.

Recent study has shown that wavefront manipulation could be insensitive to the number of unit cells in AMs [26] and although less unit cells are designed for AMs, the required phenomena can be well preserved [27-29]. For example, three-port retroreflector in an AM with six unit cells [29] was well realized by acoustic metagrating (MG) with two unit cells [27]. Therefore, acoustic binary MGs (only two unit cells) might be promising for realizing AT. By breaking spatial inversion symmetry of binary unit cells [30], AT was realized in a single-layer AM. However, by using common binary design (i.e., spatial inversion symmetry), it is almost impossible to access AT in a single-layer system. Compared with single-layer systems, dual-layer systems possess more possibilities for practical applications, as the spatial tunability makes it more flexible in manipulating wave propagation. Therefore, dual-layer MGs with common binary design are potentially a good candidate for obtaining AT, which might well solve the challenging problems of design complexities, sample fabrication, intrinsic absorption and spatial tunability in AMs.

In this work, we proposed and studied dual-layer acoustic binary MGs to realize nearly perfect AT. It is shown that due to the unique binary design, beam splitting (BS) [31] can be manipulated through adjusting the air gap of the proposed dual-layer MGs, as a result of anomalous diffraction law in phase gradient metasurfaces [26]. For a large air gap, the BS stemming from transmission channel of the first diffraction order happens for one side incidence and the total reflection from the zeroth order occurs for opposite incidence, bringing about asymmetric BS. For a closed air gap, the diffraction channels of opposite incidences concurrently are the transmission channel of the first diffraction order, leading to symmetric BS. By simply controlling the air gap, one can manipulate the diffraction channel transition, which enables the switch of asymmetric and symmetric BS. In addition, asymmetric and symmetric transmission behavior in the designed MG system can function in a certain frequency band and a wide incident range. Although the proposed MG system is dual-layer, actually it is just a simple combination of one MG, i.e., only single binary MG is used. Our work provides a simplified solution for flexibly manipulating wave propagation, enabling potential applications in acoustic communication, one-way device and noise control.

**2. Theory and model**

To clearly illustrate our idea and the concept of dual-layer MGs, let us first study wave scattering in two different single-layer binary MGs as shown in Fig. 1(a) and Fig. 1(b), where both MGs are only composed of two unit cells with $\pi$ phase difference. For wave normally impinging on MG-1 with periodic length of $p_1=2a_1<\lambda$, the transmitted/reflected wave will follow,

$$k_x^{r,t} = k_x^{in} \pm nG_1, \tag{1}$$

where $n$ is the diffraction order and $G_1=2\pi/p_1$ is reciprocal lattice vector of MG-1. Due to $G_1>k_0$ ($k_0=2\pi/\lambda$), the diffraction waves of non-zero orders are evanescent waves for normal incidence ($k_x^{in}=0$). As a result, only transmission and reflection of the $n=0$ order can happen. As MG-1 is designed with two unit cells, incident wave is totally reflected back owing to even propagation number of multiple reflections [26], leading to surface wave bounding at the transmitted surface (see Fig. 1(a)). Based on the binary unit cells in MG-1, we place them twice in a period to construct MG-2 (i.e., $p_2=2p_1$), as shown in Fig. 1(b). For wave normally incident on MG-2 with $p_2>\lambda$, the transmitted/reflected wave can take,

$$k_x^{r,t} = k_x^{in} \pm nG_2. \tag{2}$$

As $G_2=2\pi/p_2<k_0$, the diffraction orders of $n=\pm 1$ and $n=0$ are all open for $k_x^{in}=0$. However, the diffraction orders of $n=\pm 1$ can directly take one-pass propagation, which is preferential to the $n=0$ order with round-trip process [26]. Consequently, BS stemming from the transmission channel of the $n=\pm 1$ orders is realized for normal incidence, with the splitting angle depending on $p_2$, i.e., $\theta_s = \arcsin(\lambda/p_2)$. When MG-1 and MG-2 are combined together with an air gap, asymmetric BS could be achieved via asymmetric diffraction channel of opposite incidences. For positive incidence (PI), i.e., incidence along $+y$ direction, BS first occurs in the air gap, and then the divided beams from MG-2 will take the $n=\pm 1$ diffraction orders of MG-1 (see the red arrows in Fig. 1(c)). Owing to $p_2=2p_1$, perfect negative refraction can occur, leading to high-efficiency BS with the transmitted angle equal to $\theta_s$. Such propagation process could be understood from,

$$k_x^t = k_x^{in} \pm G_2 \mp G_1. \tag{3}$$

While for negative incidence (NI), i.e., incidence along $-y$ direction, incident wave is totally reflected back by taking the zeroth order as long as the air gap is larger enough to avoid surface wave coupling between MG-1 and MG-2. Therefore, low-efficiency BS can happen (see the blue arrows in Fig.1(c)). If the wavelength of incident wave is located in the region of $p_1 < \lambda < p_2$, then the splitting angle of BS is frequency-dependent and operated in a wide range, i.e., $30° < \theta_s < 90°$. However, when the air gap is closed, i.e., $\Delta = 0$, thanks to the binary design, the phase difference cross a period are "0, $\pi$, $\pi$, $2\pi$", equivalent to "0, $\pi$, $\pi$, 0". A new kind of 0-$\pi$ metagrating is generated with periodic length equal to $p_2$ and the diffraction channels are identical for opposite incidences. Consequently, high-efficiency BS from $n=\pm 1$ diffraction orders happens independent of incident directions (see Fig. 1(d)), i.e., symmetric BS. Therefore, by opening or closing the air gap, the diffraction channel of NI could be manipulated from the reflection channel of the $n=0$ order to the transmission channel of the $n=\pm 1$ order, leading to the switch between asymmetric BS and symmetric BS.

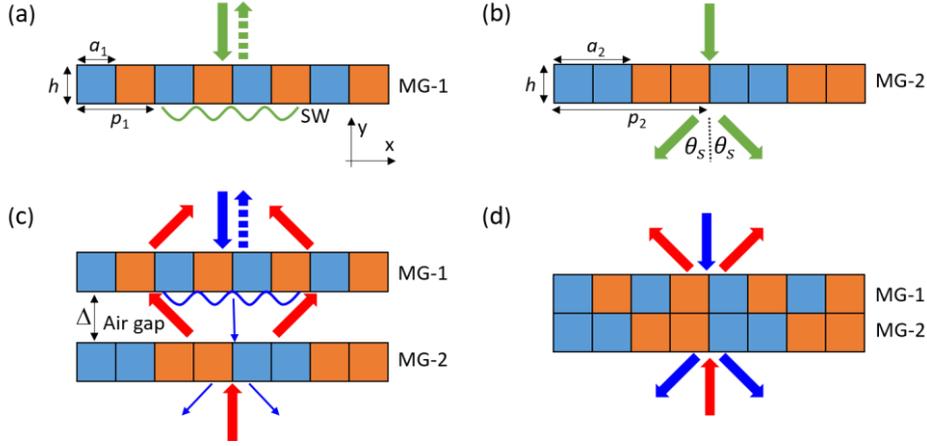

**Fig. 1.** Schematic of beam reflection in MG-1(a) and BS in MG-2 (b), in which both MGs consist of two unit cells with $\pi$ phase difference. The periodic length of MG-2 is twice of that of MG-1, i.e., $p_2 = 2p_1$. Schematic of asymmetric (c) and symmetric (d) BS in the composite structure of MG-1 and MG-2.

### 3. Metagrating design and numerical demonstration

In order to verify the above theoretical discussion, we take the splitting angle of $\theta_s = 45°$ as an example to demonstrate it. The operating wavelength is set at $\lambda = 10.0$ cm and the periodic length of MG-1 is $p_1 = \sqrt{2}/2\lambda$. The unit cell designed for the binary MGs is displayed in Fig. 2(a), where it is coiling-up space structure with height of $h = 0.5\lambda$ and width of $a_1 = p_1/2$. Six building blocks with length of $b$ are placed in the center of the unit cell and they are separated with equal distance of $d = 0.7$ cm. The thickness of all the walls made of sound-hard materials is $t = 0.1$ cm. By changing the length ($b$) of building blocks, the transmission and phase shift of the designed unit cell are shown in Fig. 2(b). To obtain binary unit cells with opposite phase shift and nearly unity transmission, the corresponding lengths of building blocks are selected as $b_1 = 0$ and $b_2 = 0.46a$ ($a = a_1 - t$), which is indicated by black dashed line in Fig. 2(b). By employing the designed binary unit cells, Fig. 2(c) and Fig. 2(d) show us the designed MG-1 and MG-2, where we can see only a single coiling-up space structure is used in the binary MGs as the other one is empty space ($b_1 = 0$). In the following, we will use the designed MGs to confirm the asymmetric and symmetric BS.

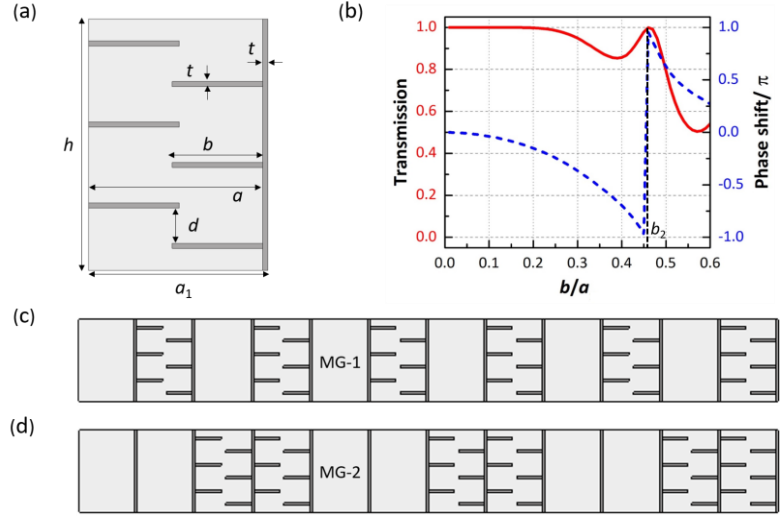

**Fig. 2.** (a) Geometric structure of unit cell designed for the binary MGs. (b) Transmission and phase shift of the unit cell vs the length of building blocks. Section of the designed (c) MG-1 and (d) MG-2, where MG-1 and MG-2 are composed of 6 and 3 supercells, respectively.

Numerical simulations using COMSOL MULTIPHYSICS are employed to check asymmetric BS. We numerically find that surface wave coupling between MG-1 and MG-2 can be greatly avoided for the air gap with $\Delta \geq 0.3\lambda$, and $\Delta = 0.5\lambda$ is chose to demonstrate the asymmetric BS as shown in Fig. 3(a) and Fig. 3(b). For the case of PI, high-efficiency BS is seen in Fig. 3(a). From the simulated field patterns, we can see that normally incident beam is firstly split in the air gap and then is crossed through MG-1, and finally the transmitted waves are two outgoing beams ($T_{\pm 1}$) with refracted angle of $\theta_s = 45°$ as expected. While for NI (see Fig. 3(b)), incident beam is almost totally reflected by MG-1, with surface wave bounding at its surface. As a result, extremely low transmitted wave of the zeroth order bumps on the MG-2, leading to quite weak BS (see Fig. 3(b)). The transmission efficiencies of BS for PI are numerically calculated as about $T_{-1} = 44.5\%$ and $T_{+1} = 52.7\%$ (the transmission coefficient of $T_{\pm 1}$ is the square of the amplitude of the diffraction coefficient of the $n = \pm 1$ order), respectively; while for NI, the transmission efficiencies of BS are $T_{-1} = 0.5\%$ and $T_{+1} = 0.6\%$ respectively. Hence, asymmetric BS is well demonstrated by numerical simulations. When the air gap is switched to $\Delta = 0$, symmetric BS is also numerically demonstrated. For PI, high-efficiency BS is observed in Fig. 3(c), where the phase difference cross a period (see the white dashed frame) is "0, $\pi$, $\pi$, $2\pi$" as expected. Similarly, high-efficiency BS for NI is also seen in Fig. 3(d). We numerically calculate the transmission efficiencies of BS as $T_{-1}=49.2\%$ and $T_{+1}=42.5\%$ for PI and $T_{-1}=55.3\%$ and $T_{+1}=36.5\%$ for NI. Therefore, by tightly tying these two MGs, high efficiency BS can be achieved regardless of incident directions.

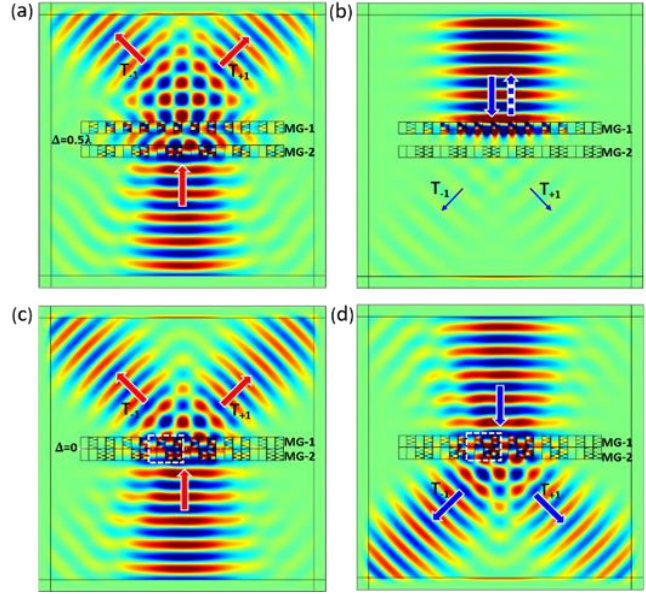

**Fig. 3.** Numerically simulated total pressure field patterns of asymmetric BS ($\Delta=0.5\lambda$) for (a) PI and (b) NI and symmetric BS ($\Delta=0$) for (c) PI and (d) NI, where MG-1 (MG-2) consists of 12 (6) supercells.

**4. Practical performance of binary metagratings**

To access the practical performance of the proposed device, the intrinsic losses stemming from the viscous and thermal dissipation should be considered in MGs. The viscous and thermal dissipation could be mimicked by numerically including losses in the air channels of MGs, with the effective parameters of air given as density of $\rho=1.21$ kg/m$^3$ and sound speed of $c=343(1+\gamma i)$ with $\gamma>0$. Under such equivalent operation for the intrinsic losses, the simulated results will be closer to experimental measurements, referring to the experimental works [32]. Accordingly, simulations for asymmetric BS are performed by considering practical loss of $\gamma=0.028$ in MGs, which is extracted from an experiment work related to the coiling-up space structure [32]. As shown in Fig. 4(a) and Fig. 4(b), good performance of asymmetric BS is well preserved, and the transmission efficiencies of PI and NI are 52.3% and 2.6%, respectively, due to the intrinsic dissipation in MGs. Simulated results are also performed for symmetric BS, and as the corresponding results are similar with that in Fig. 4(a), they are not displayed here. Since the air channel of our designed structure is much wider than that in Ref. [32], the practical loss might be smaller (e.g., $\gamma=0.01$) and the performance of the device will be further promoted.

In addition, to quantitatively study the performance of asymmetric/symmetric transmission behavior, the contrast ratio of transmission energy is analyzed as a function of frequency. The contrast ratio of transmission energy is defined as $\eta = |E_P - E_N|/(E_P + E_N)$, where $E_P$ and $E_N$ are transmission energy for PI and NI, respectively. The transmission energy is tested at the surface away from MG-1 or MG-2 with distance of $\lambda$ and the transmission energy is proportional to the square of the field amplitude. The contrast ratio vs frequency for $\Delta=0.5\lambda$ (AT) and $\Delta=0$ (symmetric transmission) is shown in Fig. 4(c), where the frequencies of interest are considered near the operating frequency (3430 Hz), i.e., from 3000 to 3800 Hz. With the increase of loss, the contrast ratio is gradually decreased and good performance ($\eta \geq 0.8$) of AT can function approximately from 3300 to 3600 Hz for each case. While for the symmetric transmission, the value of loss almost doesn't affect the contrast ratio and good performance ($\eta \leq 0.2$) of symmetric transmission can work in the whole range. Therefore, even considering practical losses in the designed MG system, asymmetric/symmetric transmission with good performance still can happen.

In fact, the asymmetric/symmetric transmission not only happens at normal incidence but also occurs at other incidences. To clearly reveal this, we numerically show the contrast ratio vs incident angle for different losses at 3430 Hz in Fig. 4(d), where only half incident range ($0° \leq \theta \leq 70°$) is considered owing to symmetrically angular resonance in the binary MG system. AT with good performance is found in two incident regions, i.e., $0° \leq \theta \leq 17.8°$ and $40° \leq \theta \leq 50°$ near $\theta_s=45°$, which are separated by the critical angle of $\theta_c = \arcsin(\lambda/p_1 - 1) \approx 24.5°$. For the incidence within $\theta_c$, strong transmission energy happens for PI and weak transmission energy occurs for NI. For $0° \leq \theta \leq 17.8°$, with the increase of incident angle, transmission energy of PI (NI) gradually decreases (increases). Therefore, the contrast ratio reduces with the increase of incident angle. While for the incidence beyond $\theta_c$, the high contrast ratio in $40° \leq \theta \leq 50°$ is caused by high transmission of NI and low transmission of PI. For example, when incident beam with $\theta=45°$ bumps on MG-2 (PI), it will normally transmit through MG-2 and then the transmitted beam is totally reflected back by MG-1, leading to low transmission (see the red arrows in the insert of Fig. 4(d)). While for beam with $\theta=45°$ incident on MG-1 (NI), the transmitted beam can pass through MG-1 with $\theta=-45°$ and it is normally bended by MG-2, bringing about high transmission (see the blue arrows in the insert of Fig. 4(d)). With the increase of loss, the contrast ratio of AT in $0° \leq \theta \leq 17.8°$ gradually decreases and it near $\theta_s=45°$ is gradually raised, with the incident range broaden. It is caused by further reduction of transmission energy of PI when loss is increased. Furthermore, symmetric transmission can almost cover the whole incident range, as shown in Fig. 4(d). Based on Fig. 4, we can find asymmetric/symmetric transmission in the binary MG system can function over a certain range of frequency near the targeted one, with wide angle response observed.

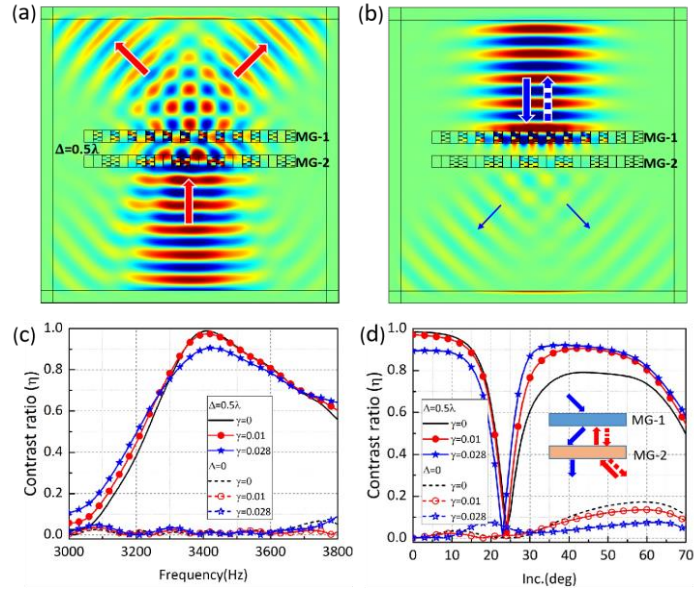

**Fig. 4.** Numerically simulated field patterns of asymmetric BS for (a) PI and (b) NI, where a practical loss of $\gamma=0.028$ from experiment is considered in the MGs. Contrast ratio of transmission energy vs (c) frequency and (d) incident angle for AT ($\Delta=0.5\lambda$) and symmetric transmission ($\Delta=0$), where the black solid (dashed) curve, red solid (dashed) curve with circles and blue solid (dashed) curve with stars denote the related cases of lossless ($\gamma=0$), lossy ($\gamma=0.01$) and lossy ($\gamma=0.028$) MGs.

## 5. Discussion and conclusion

In conclusion, by designing simple binary unit cells only made of a single coiling-up space structure, we have numerically demonstrated asymmetric BS in dual-layer MGs, which can response in a certain frequency band and a wide incident angle. By closing the air gap of dual-layer MGs, BS can be switched from AT to symmetric transmission. The splitting angle could be theoretically tuned from $\theta_s = 30°$ to $\theta_s = 90°$, but the transmitted energy of PI will gradually reduce with the increase of $\theta_s$ due to the intrinsic impedance mismatching [33]. Although the thickness of the unit cell is designed with half wavelength to demonstrate BS in the binary MGs, it can be further reduced with sub-wavelength thickness [34]. Our proposed scheme to obtain AT in a neat and practical way could be extended to other physical systems, such as electromagnetic waves, elastic waves [35] and water waves, leading to more functional devices with simplified design.

## Acknowledgements

This work is supported by National Natural Science Foundation of China (11904169, 11974010, 61675095 and 11604229) and Natural Science Foundation of Jiangsu Province (BK20190383 and BK20171206).